\documentclass[journal,10pt,twocolumn]{IEEEtran}
\usepackage{epsfig}
\usepackage{graphicx}
\usepackage{subfigure}

\usepackage{amssymb}
\usepackage{makeidx}
\usepackage{amsmath}
\usepackage{graphicx}
\usepackage{epsf}
\usepackage{epsfig}
\usepackage{psfig}
\usepackage{ccaption}
\usepackage{array}
\usepackage{tabularx}
\usepackage{multirow}
\usepackage{epsfig}
\usepackage{cite}
\usepackage{pifont}

\begin{document}

\title{Decoding the `Nature Encoded' Messages for Distributed Energy Generation Control in Microgrid}
\author{Shuping Gong, Husheng Li, Lifeng Lai and Robert. C. Qiu
\thanks{S. Gong and H. Li are with the Department of Electrical Engineering and Computer Science, the University of Tennessee, Knoxville, TN (email:
sgong1,husheng@eecs.utk.edu). L. Lai is with the Department of Systems Engineering, University of Arkansas, Little Rock, AR, (email: lxlai@ualr.edu). R. C. Qiu is with the Department of Electrical and Computer Engineering, Tennessee Technological University, Cookeville, TN, (email: rqiu@ttn.edu). This work was supported by the National
Science Foundation under grants CCF-0830451 and ECCS-0901425.}}

\maketitle

\begin{abstract}
The communication for the control of distributed energy generation (DEG) in microgrid is discussed. Due to the requirement of realtime transmission, weak or no explicit channel coding is used for the message of system state. To protect the reliability of the uncoded or weakly encoded messages, the system dynamics are considered as a `nature encoding' similar to convolution code, due to its redundancy in time. For systems with or without explicit channel coding, two decoding procedures based on Kalman filtering and Pearl's Belief Propagation, in a similar manner to Turbo processing in traditional data communication systems, are proposed. Numerical simulations have demonstrated the validity of the schemes, using a linear model of electric generator dynamic system.
\end{abstract}

\section{Introduction}
In recent years, smart grid has attracted significant studies due to its capability of improving the efficiency and robustness of power grid \cite{ISO2009}. An important component in smart grid is the microgrid \cite{Hatziargyrion2007}\cite{Piagi2006}, which is a network containing multiple distributed energy generators (DEGs, like solar panels, microturbines or wind turbines) and multiple loads. A microgrid can either be connected to or be separated from the power grid. In normal situations, the microgrid is connected to the power grid and obtains energy according to its own energy generation. When the power grid experiences an emergency, e.g., a large area blackout, the microgrid can be disconnected from the power grid and work in an island mode. At this time, the power load within the microgrid will be supported by the DEGs. Such a microgrid has been implemented in many testbeds, e.g., the Electric Reliability Technology Solutions (CERTS) Microgrid supported by the US Department of Energy.

An important aspect of the microgrid is the communication infrastructure for controlling the DEGs since they could be located at different places. In such a communication infrastructure, the observations at sensors are sent to a processor (either centralized or decentralized) for system state monitoring or control. In traditional microgrids, wired communications are employed. For example, an RS-485 (digital computer based) link or an ethernet is used in the Oak Ridge National Lab (ORNL) microgrid system \cite{Kueck2003}. An alternative approach for the communication for DEG control is to use wireless communication systems due to its fast deployment. In wireless medium, fading, noise and interference may cause damage to the received signal of messages for DEG control. Therefore, a channel coding scheme is needed to protect the messages. However, there is a stringent requirement on the delay of messages due to the dynamics of DEGs. Hence, it is impossible to accumulate sufficiently many bits for a long codeword length and the corresponding coding gain since this will incur a substantial delay. Therefore, an effective approach is to use short codeword length which loses the protection on the bits in the message.

\begin{figure}
  \centering
  \includegraphics[scale=0.4]{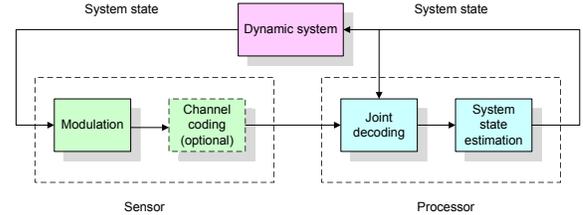}
  \caption{An illustration of the communication procedure and dynamic system.}\label{fig:illu}
\end{figure}

In this paper, we study {\em the decoding procedure (including the demodulation procedure) for the messages with no or weak channel coding protection in the context of DEG control in microgrid.} The key point is to {\em exploit the redundancy in the system state, which can be considered as a `nature encoding' for the messages}. In this paper, a linear system model is adopted, where $\mathbf{x}(t+1)=\mathbf{A}\mathbf{x}(t)+\mathbf{B}\mathbf{u}(t)+\mathbf{n}(t)$ is used to describe the dynamics of system state $\mathbf{x}$ subject to the control $\mathbf{u}(t)$ and noise $\mathbf{n}(t)$. We observe that the system state is similar to the convolutional codes except that the `encoding' of the system state is in the real field instead of the Galois field. Hence, the message actually has been channel coded by the nature although no explicit or little explicit channel coding is used at the transmitter. We use both schemes of Kalman filtering and Pearl's Belief Propagation (BP) for the soft decoding, combined with the soft demodulation to improve the reliability of demodulation and decoding. A practical dynamic system for the DEGs will be used for numerical simulation, which demonstrates the performance gain of incorporating the inherent redundancy in the `nature encoding'. The procedure is illustrated in Fig. \ref{fig:illu}. Note that such a scheme is similar to the {\em Turbo processing techniques} like Turbo decoding \cite{Berrou1993}, Turbo multiuser detection \cite{Alexander1999} and Turbo equalization \cite{Tuchler2002} in wireless communication systems. However, the `nature encoding' is analog and implicit, thus resulting in a different processing procedure.

The remainder of this paper is organized as follows. The system model is introduced in Section \ref{sec:system}. The decoding procedure is discussed for the schemes of Kalman filtering based heuristic approach and the BP approach in Sections \ref{sec:Kalman} and \ref{sec:BP}, respectively. Numerical simulation results are shown in Section \ref{sec:numerical}. Finally, the conclusions are drawn in Section \ref{sec:conclusion}.

\section{System Model}\label{sec:system}

\subsection{Linear System}
We consider a discrete time linear dynamic system, whose system state evolution is given by
\begin{eqnarray} \label{eqn:sysModel}
\left\{
\begin{array}{ll}
\mathbf{x}(t+1)=\mathbf{A}\mathbf{x}(t)+\mathbf{B}\mathbf{u}(t)+\mathbf{n}(t),\\
\mathbf{y}(t)=\mathbf{C}\mathbf{x}(t)+\mathbf{w}(t)
\end{array}
\right.,
\end{eqnarray}
where $\mathbf{x}(t)$ is the $N$-dimensional vector of system state at time slot $t$, $\mathbf{u}(t)$ is the $M$-dimensional control vector, $\mathbf{y}$ is the $K$-dimensional observation vector and $\mathbf{n}$ and $\mathbf{w}$ are noise vectors, which are assumed to be Gaussian distributed with zero expectation and covariance matrices $\mathbf{\Sigma}_n$ and $\mathbf{\Sigma}_w$, respectively. For simplicity, in this paper we do not consider $\mathbf{u}(t)$.

We assume that the observation vector $\mathbf{y}(t)$ is obtained by a sensor\footnote{It is easy to extend to the case of multiple sensors.}. The sensor quantizes each dimension of the observation using $B$ bits, thus forming a bit sequence which is given by
\begin{eqnarray}
\mathbf{b}(t)=\left(b_1(t),b_2(t),...,b_{KB}(t)\right).
\end{eqnarray}

\subsection{Communication System}
Suppose that binary phase shift keying (BPSK) is used for the transmission from the sensor to the controller. The bit sequence is passed through an optional channel encoder, which generates an $L$-bit sequence $\mathbf{s}(t)$. Then, the received signal at the controller is given by
\begin{eqnarray}
\mathbf{r}(t)=\mathbf{s}(t)+\mathbf{e}(t),
\end{eqnarray}
where the additive white Gaussian noise $\mathbf{e}(t)$ has a zero expectation and variance $\sigma_e^2$. Note that we ignore the fading and normalize the transmit power to be 1. The algorithm and conclusion in this paper can be easily extended to the case with different types of fading.

\section{Kalman Filtering based Heuristic Approach}\label{sec:Kalman}
In this section, we adopt a heuristic approach, which is based on Kalman filtering, to exploit the redundancy in the system state. We first carry out the Kalman filtering and then apply the prediction to the soft demodulation.

\subsection{Kalman Filtering}
When the observations $\mathbf{y}(t)$ are sent to the controller perfectly, the controller can use the following Kalman filtering to predict the future system state, whose expectation is given by
\begin{eqnarray}\label{eqn:kfMuPredict}
\mathbf{x}(t+1|t)=\mathbf{A}\mathbf{x}(t|t),
\end{eqnarray}
where
\begin{eqnarray}
\mathbf{x}(t|t)=\mathbf{x}(t|t-1)+\mathbf{K}(t)\left(\mathbf{y}-\mathbf{C}\mathbf{x}(t|t-1)\right),
\end{eqnarray}
and
\begin{eqnarray}
\mathbf{K}(t)=\mathbf{\Sigma}(t|t-1)\mathbf{C}^T\left(\mathbf{C}\mathbf{\Sigma(t|t-1)\left(\mathbf{C}\right)^T+\mathbf{\Sigma}_o}\right)^{-1},
\end{eqnarray}
and the covariance matrix given by
\begin{eqnarray}
\mathbf{\Sigma(t|t)}=\mathbf{\Sigma}(t|t-1)-\mathbf{K}_t\mathbf{C}\mathbf{\Sigma}(t|t-1),
\end{eqnarray}
where
\begin{eqnarray}\label{eqn:kfCovPredict}
\mathbf{\Sigma(t+1|t)}=\mathbf{A}\mathbf{\Sigma(t|t)}\mathbf{A}^T+\mathbf{\Sigma}_p.
\end{eqnarray}

\subsection{Soft Decoding and Demodulation}\label{sec:softDecodingDemod}
Based on the Kalman filtering, the controller can obtain the distribution of the observation, which is Gaussian distributed with the expectation $\mathbf{x}(t|t-1)$ given by Eq. (\ref{eqn:kfMuPredict}) and the covariance $\mathbf{\Sigma}(t|t-1)$ given by Eq. (\ref{eqn:kfCovPredict}).

Because different dimensions in the observation $\mathbf{y}_{t}$ are not independent, it is challenging to directly compute the {\em a priori} probability for each bit, which is given by (suppose that $i$ is used to describe the $i$-th bit in  $\mathbf{b}(t)$)
\begin{eqnarray}
\xi_i(t)&\triangleq& P(b_i(t)=1|\mathbf{y}(-\infty:t))\nonumber\\
           &=&\frac{1}{\sqrt{2\pi |\mathbf{\Sigma}(t|t-1)|}}\int I(b_i(t)=1, \mathbf{y}(t)=\mathbf{y})\nonumber\\
           &\times&\exp(-\frac{1}{2}(\mathbf{y}-\mathbf{x}(t|t-1))^T \times \mathbf{\Sigma}^{-1}(t|t-1) \nonumber \\
&\times&(\mathbf{y}-\mathbf{x}(t|t-1)))d\mathbf{y}.
\end{eqnarray}

We propose to use the Monte Carlo simulation to obtain a series of samples of $\left\{\mathbf{y}_{t}\right\}$ based on the prediction of Kalman filtering, then quantize these samples to obtain a series of samples $\mathbf{b}_{t-1}$ and calculate the prior probability $\xi_i(t)$ for $\mathbf{b}_{t}$.
We use $\left\{\xi_i(t)\right\}$ as the {\em a priori} probability of being 1 for demodulating $\mathbf{b}(t)$. Then, the {\em a posteriori} probability of $b_i(t)$ is given by
\begin{small}
\begin{eqnarray}
&&P(b_i(t)=1|\mathbf{r}(-\infty:t))\nonumber\\
&=&\frac{P(\mathbf{r}(t)|b_i(t)=1)\xi_i(t)}{P(\mathbf{r}(t)|b_i(t)=1)\xi_i(t)+P(\mathbf{r}(t)|b_i(t)=0)(1-\xi_i(t))},
\end{eqnarray}
\end{small}
where
\begin{eqnarray}
P(\mathbf{y}(t)|b_i(t)=1)=\frac{1}{\sqrt{2\pi \sigma_0^2}}\exp\left(-\frac{(r_i(t)-1)^2}{2\sigma_e^2}\right). \\
P(\mathbf{y}(t)|b_i(t)=0)=\frac{1}{\sqrt{2\pi \sigma_0^2}}\exp\left(-\frac{(r_i(t) + 1)^2}{2\sigma_e^2}\right).
\end{eqnarray}

Note that the Kalman filtering is no longer rigorous in the networked control system due to the quantization error and possible decoding error. The proposed heuristic approach is based on the assumption that the Kalman filtering is very precise. As will be shown in the numerical results, this approach will be seriously affected by the propagation of decoding errors.

\section{BP based Iterative Decoding}\label{sec:BP}
\begin{figure}
  \centering
  \includegraphics[scale=0.55]{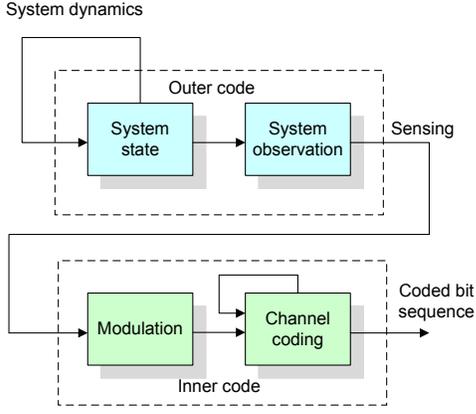}
  \caption{An illustration of the coding structure.}\label{fig:coding}
\end{figure}

In this section, we consider the iterative decoding using BP with the mechanism of message passing between the system state and received signals. The key observation is that {\em the entire information passing is similar to a concatenated coding structure}, as illustrated in Fig. \ref{fig:coding}. The outer coding is carried out by the dynamic system, where the system state $\mathbf{x}$ is encoded similarly to a convolutional code and the observation $\mathbf{y}$ is linearly encoded by the observation matrix $\mathbf{C}$ and the system state $\mathbf{x}$. The inner code is the explicit encoding of the observation vector. Hence, we can adopt the iterative decoding approach in Turbo codes or LDPC codes \cite{Richardson2001}. In a sharp contrast, the proposed Kalman filtering based heuristic approach has only one round.

\cite{McEliece1998} has shown that many iterative decoding algorithm, such as Turbo decoding can be considered as the application of Pearl's BP algorithm \cite{Pearl1988}. In this section, we first illustrate the principle through an example and then explain how to apply Pearl's BP into our dynamic state system.

\subsection{Introduction of Pearl's BP}
\begin{figure}
  \centering
  \includegraphics[scale=0.4]{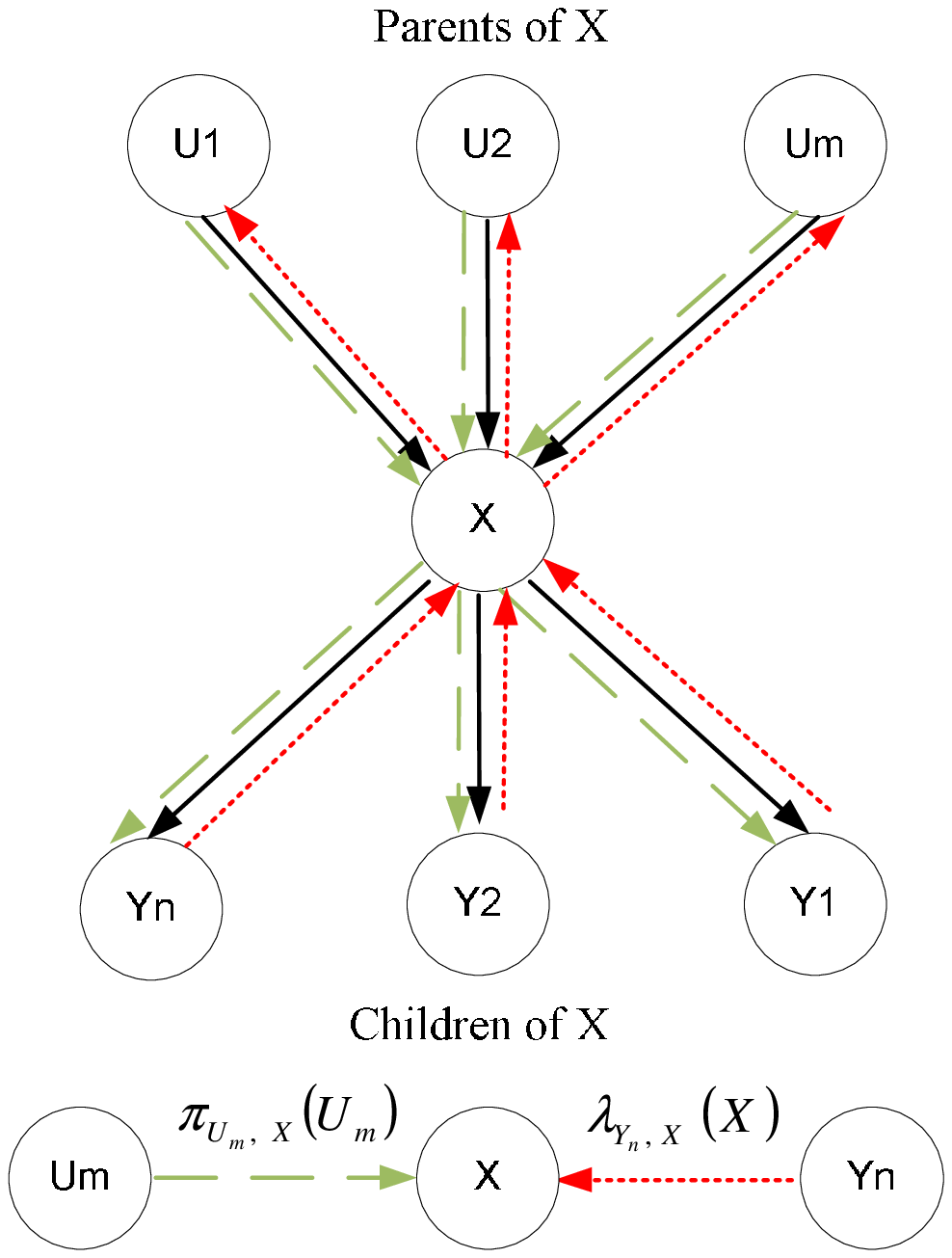}
  \caption{Message Passing of BP.}\label{fig:gbp}
\end{figure}
As shown in Fig. \ref{fig:gbp}, random variable $X$ has parents $U_1, U_2, \cdots, U_M$ and children $Y_1, Y_2, \cdots, Y_N$. The message passing of Pearl's BP is indicated by green arrows and red arrows in the figure. Green arrows transmit $\pi$-message which is sent from parent to its children. For instance, the message passing from $U_m$ to $X$ is $\pi_{U_m, X}(U_m)$, which is the prior information of $U_m$ conditioned on all the information $U_m$ has received. Red arrows transmit $\lambda$-message which is from children to its parent. For instance, the message passing from $Y_n$ to $X$ is $\lambda_{Y_n,X}(X)$, which is the likelihood of X based on the information $Y_n$ has received. After $X$ receives all $\pi$-message $\pi_{U_m, X}(U_m)$  from its parents $U_1, U_2, \cdots, U_M$ and  all $\lambda$-message $\lambda_{Y_n,X}(X)$, $X$ from its children $Y_1, Y_2, \cdots, Y_N$, $X$ updates its belief information $BEL_X(\mathbf{x})$ and transmits $\lambda$-messages $\lambda_{X,U_m}(U_m)$ to its parents and $\pi$-message $\pi_{X, Y_n}(\mathbf{x})$ to its children. The expressions of the quantities are given by
\begin{eqnarray}
\pi_X(\mathbf{x}) &=& \sum_{\mathbf{U}}{p(\mathbf{x}|\mathbf{U})\prod_{m=1}^M{\pi_{U_m, X}(U_m)}} \\
\gamma_X(\mathbf{U}) &=& \sum_{\mathbf{x}}{\prod_{n=1}^N{\lambda_{Y_n,X}(\mathbf{x})}p(\mathbf{x}|\mathbf{U})} \\
BEL_X(\mathbf{x}) &=& \alpha \times \prod_{n=1}^N{\lambda_{Y_n,X}(\mathbf{x}) \times \pi_X(\mathbf{x})} \\
\lambda_{X,U_m}(U_m) &=& \sum_{\mathbf{U}, \neq U_m }{\gamma_X(\mathbf{U}) \times \prod_{j \neq m}{\pi_{U_j, X}(U_j)}}  \\
\pi_{X, Y_n}(\mathbf{x})
&=& \pi_X(\mathbf{x}) \times \prod_{i \neq n}{\lambda_{Y_i, X}(\mathbf{x})}
\end{eqnarray}
where $\mathbf{U} = (U_1, U_2, \cdots, U_M)$ and $\mathbf{Y} = (Y_1, Y_2, \cdots, U_N)$

In the initialization procedure of Pearl's BP, some initial values are needed for the running of Pearl's BP. The initial values are defined as
\begin{eqnarray}\label{eqn:bpInitialLambda}
\lambda_{X,U}(\mathbf{u}) &=&  \left\{
\begin{array}{ll}
p(\mathbf{x}_0|\mathbf{u}), & X \text{is evidence}, \mathbf{x} = \mathbf{x}_0 \\
 1, & X \text{is not evidence}
\end{array} \right.,
\end{eqnarray}
and
\begin{eqnarray}\label{eqn:bpInitialPi}
\pi_{X,Y}(\mathbf{x}) &=&  \left\{
\begin{array}{ll}
\delta(\mathbf{x}, \mathbf{x}_0), & X \text{is evidence}, \mathbf{x} = \mathbf{x}_0 \\
 p(\mathbf{x}),  & X \text{ is source, not evidence}
\end{array} \right..
\end{eqnarray}

\subsection{Application of Pearl's BP in Microgrid}
\begin{figure}
  \centering
  \includegraphics[scale=0.5]{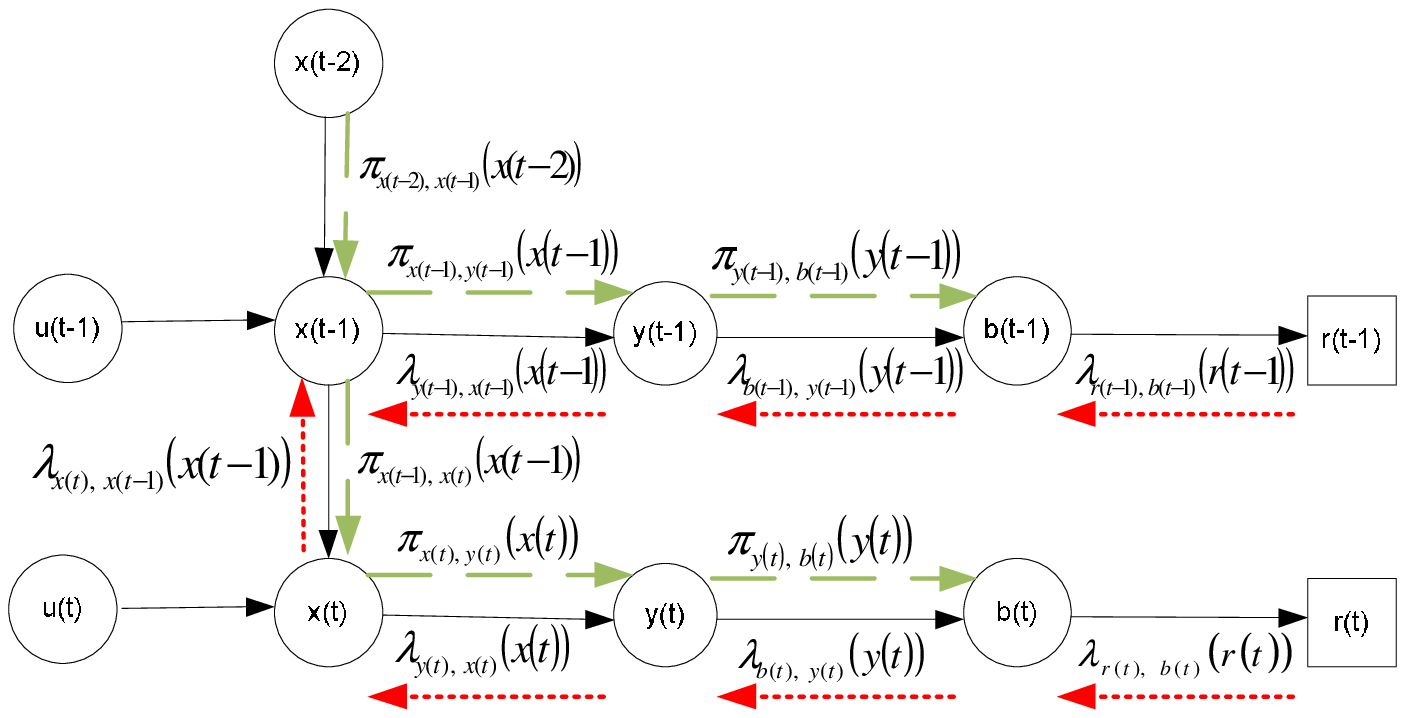}
  \caption{Bayesian network structure and message passing for the dynamic system.}\label{fig:bp}
\end{figure}

The Bayesian network structure of the control system and communication system in the DEG networked control in microgrid is shown for three time slots in Fig. \ref{fig:bp}. In the Bayesian network, the system state is dependent on the previous system state and the control action; the observation is dependent on the system state; the uncoded bits are dependent on the observation vector; the received signal is dependent on the uncoded bits. Here we omit the coded bits as the relationship between the uncoded bits and the coded bits is deterministic.
Fig. \ref{fig:bp} shows the Bayesian Network structure for the dynamic system with three observations: $\mathbf{x}_{t-2}$, $\mathbf{x}_{t-1}$ and $\mathbf{x}_{t}$.

Based on the Bayesian network structure, the iterative decoding procedure can be derived. Fig \ref{fig:bp} shows the message passing in the dynamic system. $\mathbf{x}_{t-2}$ summarizes all the information obtained from previous time slots and transmits $\pi$-message $\pi_{\mathbf{x}_{t-2},\mathbf{x}_{t-1}}(\mathbf{x}_{t-2})$ to $\mathbf{x}_{t-1}$. The BP procedure can be implemented in synchronous or asynchronous manners. As the decoding process has a large overhead, we implement asynchronous Pearl's BP. The updating order and message passing in one iteration is as follows: step 1): $\mathbf{x}_{t-1}$ \ding{213} $\mathbf{y}_{t-1}$; step 2): $\mathbf{y}_{t-1}$ \ding{213} $\mathbf{b}_{t-1}$; step 3): $\mathbf{b}_{t-1}$ \ding{213} $\mathbf{y}_{t-1}$; step 4): $\mathbf{y}_{t-1}$ \ding{213} $\mathbf{x}_{t-1}$; step 5): $\mathbf{x}_{t-1}$ \ding{213} $\mathbf{x}_{t}$; step 6): $\mathbf{x}_{t}$ \ding{213} $\mathbf{y}_{t}$; step 7): $\mathbf{y}_{t}$ \ding{213} $\mathbf{b}_{t}$; step 8): $\mathbf{b}_{t}$ \ding{213} $\mathbf{y}_{t}$; step 9): $\mathbf{y}_{t}$ \ding{213} $\mathbf{x}_{t}$; step 10): $\mathbf{x}_{t}$ \ding{213} $\mathbf{x}_{t-1}$; step 11): $\mathbf{x}_{t-1}$ updates information. The mathematical derivation for each step is provided in Appendix \ref{appdx:BP}. Due to the limited space, we provide only the most challenging steps. The full derivation will be given in our journal version.

\section{Numerical Simulations}\label{sec:numerical}
In this section, we use numerical simulations to demonstrate the algorithms proposed in this paper.

\subsection{Dynamic System Model}
We consider a dynamic system of DEG in which the system state is a 7-dimensional vector. The system is in the continuous time. The system dynamics is described by a differential equation $\dot{\mathbf{x}}(t)=\mathbf{A}'\mathbf{x}(t)$, where the matrix $\mathbf{A}'$ is given by (\ref{eq:matA}) (Example 12.9 in \cite{Machowski2008}). The physical meanings of the system states are given in Table \ref{tab:meaning}. For simplicity, we assume that the system is unregulated, i.e., $\mathbf{B}=0$, and the sensor can sense the system state directly, i.e., $\mathbf{C}=\mathbf{I}$. We approximate the continuous time system using the discrete time system with a small step size $\delta t$. Therefore, the matrix $\mathbf{A}$ in the discrete time system is given by $\mathbf{A}=\mathbf{I}+\delta t\mathbf{A}'$.

\begin{table}[ht]
\caption{Physical Meaning of System States}
\label{tab:meaning}\centering
\begin{tabular}{c c}
  \hline
    $x_1$,$x_2$ & rotor swings \\
  \hline
    $x_3$ & excitation circuit \\
  \hline
    $x_4$ & damping circuit in the $d$-axis and excitation circuit  \\
  \hline
    $x_5$ & damping circuit in the $q$-axis \\
  \hline
    $x_6$ & voltage controller and excitation circuit \\
  \hline
    $x_7$ & voltage controller \\
  \hline
\end{tabular}
\end{table}

We run the simulations using Matlab to compare the performances of Kalman filtering and Pearl's BP based algorithms for systems with and without channel coding. The baseline approach is the separated Kalman filtering and decoding process. In the following, these three algorithms are referred as `KF with Prior', `Pearl BP' and `KF', respectively. The performance metrics are the mean square error (MSE) of each sample and the average bit error rate (BER). Each simulation runs 1000 times slots. The configuration for both systems are as follows: for the dynamic system of DEG described in (\ref{eqn:sysModel}): $\delta t = 0.01$, $\mathbf{\Sigma}_p = 0.05$, $\mathbf{\Sigma}_0 = 0.05$. Each dimension of the observation $\mathbf{y}_t$ is quantized with 16 bits, and the dynamic range for quantization is $[-200, 200]$. A 1/2 rate Recursive Systematic Convolutional (RSC) code is used as the channel coding scheme; and the code generator is $g = [1, 1, 1]$. The decoding algorithm is Log-Map algorithm.

\newcounter{mytmpeqncnt}
\begin{figure*}[!t]
\normalsize \setcounter{mytmpeqncnt}{\value{equation}}
\setcounter{equation}{19}
\begin{eqnarray}\label{eq:matA}
\mathbf{A}'=\left(
             \begin{array}{ccccccc}
               0 & 1 & 0 & 0 & 0 & 0 & 0 \\
               -20.316 & 0 & 0 & -25.048 & -1.411 & 0 & 0 \\
               -0.061 & 0 & -0.773 & -0.083 & 0.018 & 15.06 & 30.12 \\
               -0.213 & 0 & 7.050 & -5.026 & 0.063 & 0 & 0 \\
               -2.654 & 0 & 0 & -1.463 & -12.958 & 0 & 0 \\
               0 & 0 & 0 & 0 & 0 & 0 & 1 \\
               -0.008 & 0 & 0 & -0.565 & 0.971 & -3.33 & -33.33 \\
             \end{array}
           \right)
\end{eqnarray}
\hrulefill \vspace*{4pt}
\end{figure*}
\setcounter{equation}{20}

\subsection{Uncoded Case}

\begin{figure}
  \centering
  \includegraphics[scale=0.4]{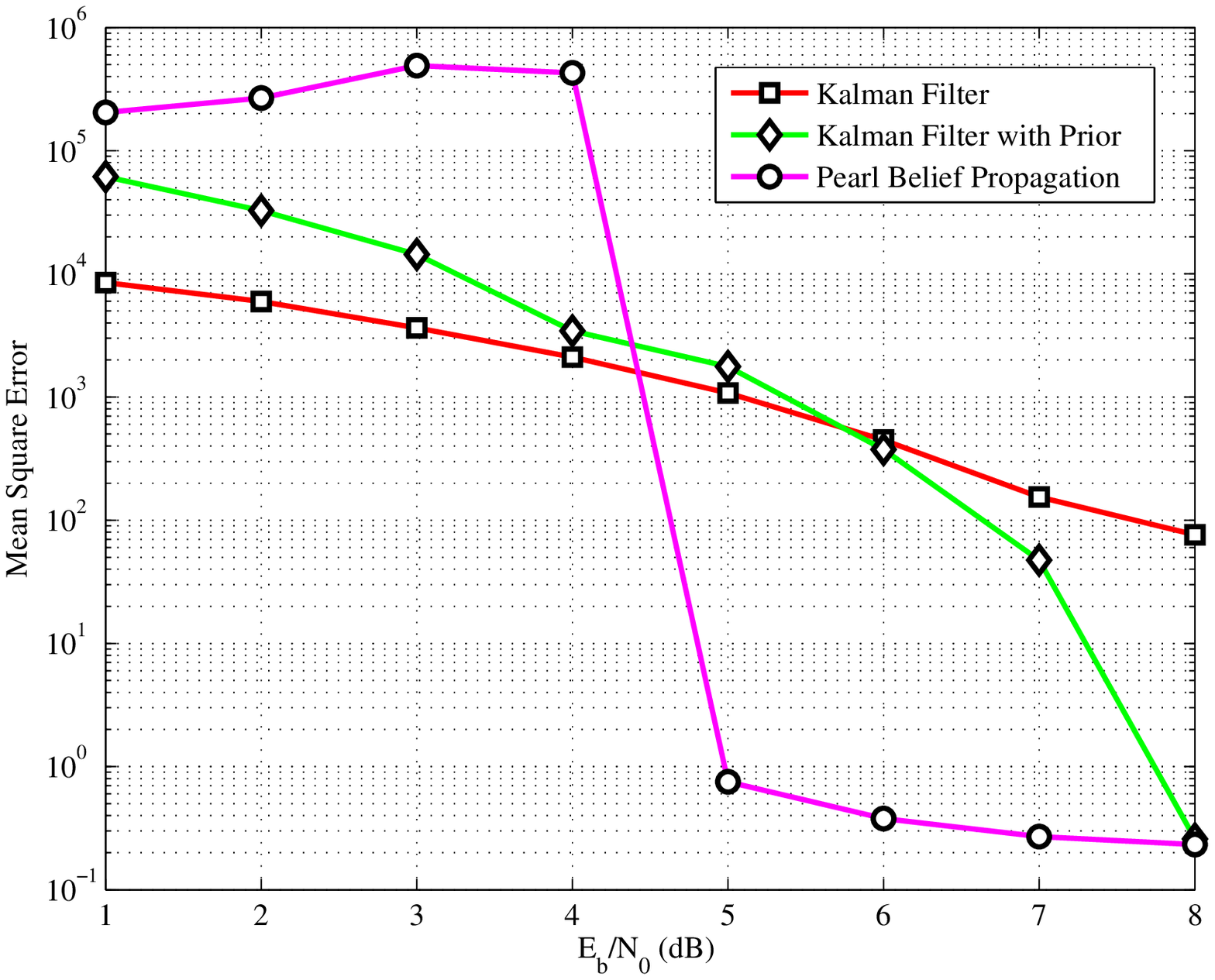}
  \caption{Mean Square Error comparison for system without channel coding}\label{fig:uncoded_mse}
\end{figure}

\begin{figure}
  \centering
  \includegraphics[scale=0.4]{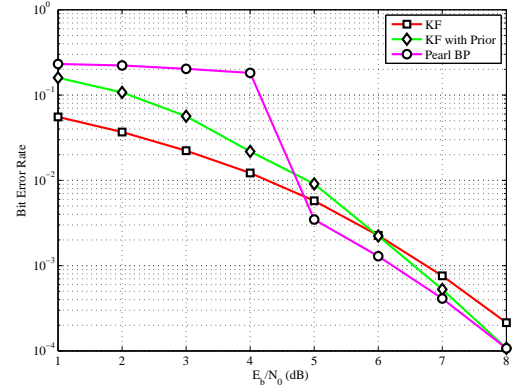}
  \caption{Bit Error Rate comparison for system without channel coding}\label{fig:uncoded_ber}
\end{figure}

Figures \ref{fig:uncoded_mse} and \ref{fig:uncoded_ber} show the simulation results for the uncoded case with different $\frac{E_b}{N_0}$. When $\frac{E_b}{N_0} \geq 5$, the performance of `Pearl BP' is much better than `KF' and `KF with Prior'; when $\frac{E_b}{N_0} \geq 7$, the performance of `KF with Prior' is better than `KF'. This demonstrates that the redundancy in the system state can improve the performance when $\frac{E_b}{N_0}$ is high. Compared with `KF with Prior', the performance gain of `Pearl BP' stems from the utilization of soft decoding output in the system state estimation. However, in the low $\frac{E_b}{N_0}$ regime, the performances of `Pearl BP' and `KF with Prior' are worse than 'KF'. This is because the cascading effect of the decoding error and system sate estimation error. The worst case is that some highest bits of observations are not decoded correctly, then the estimated observation largely deviates from the correct value; and then this error will propagate to succeeding time slots and thus will lead to the collapse of `KF with Prior' or `Pearl BP' algorithm.

\subsection{Coded Case}

\begin{figure}
  \centering
  \includegraphics[scale=0.4]{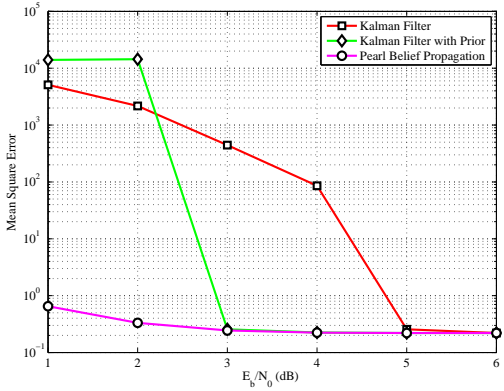}
  \caption{Mean Square Error comparison for system with channel coding}\label{fig:coded_mse}
\end{figure}

\begin{figure}
  \centering
  \includegraphics[scale=0.4]{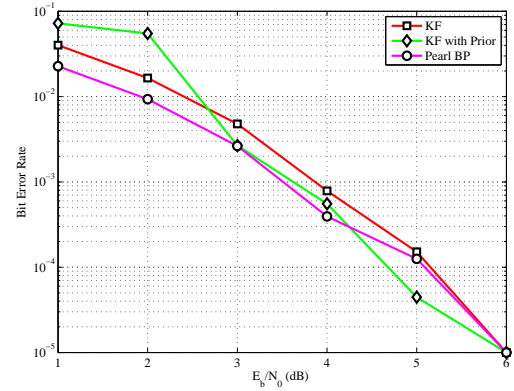}
  \caption{Bit Error Rate comparison for system with channel coding}\label{fig:coded_ber}
\end{figure}

Figures \ref{fig:coded_mse} and \ref{fig:coded_ber} show the simulation results for the channel coded case with different $\frac{E_b}{N_0}$. As shown in the figures, the performance of `Pearl BP' is always the best except that the bit error rate is worse than `KF with Prior' occasionally. In a contrast to the `Pearl BP' for the uncoded case, `Pearl BP' does not collapse for coded case due to the gain of channel coding. When $\frac{E_b}{N_0} \geq 3$, the performance of `KF with Prior' is better than that of 'KF'. These results demonstrate that using the redundancy in system state can improve performance in the coded case. The performance gain of `Pearl BP' over `Kalman Filter with Prior' in low $\frac{E_b}{N_0}$ shows the gain of using soft decoding output in state estimation. When $\frac{E_b}{N_0} \geq 5$, these three algorithms have good performance which is due to the gain of channel coding.

\section{Conclusions}\label{sec:conclusion}
In this paper, we have proposed to use Kalman Filter based and Pearl's BP based decoding procedure (including the demodulation procedure) to exploit the redundancy, i.e., the nature encoding in the system state for the system with no or weak channel coding protection in the context of DEG control in microgrid. The numerical simulation results have shown that, when $\frac{E_b}{N_0}$ is high, the proposed algorithms achieve significant performance gain, compared with traditional separated decoding and system state estimation. However, when $\frac{E_b}{N_0}$ is low, the proposed algorithms may incur performance degradation due to the error propagation. This motivates us to study the mechanism to prevent such an adversarial effect.

\appendices

\section{Derivation of Key Steps in the Pearl's BP}\label{appdx:BP}
In the following derivation part, we will use the following two basic equations, where $\mathcal{N}(x,\mu,\sigma$ is the probability density function of Gaussian distribution with expectation $\mu$ and variance $\sigma$ at $x$:
\begin{eqnarray}
&&\int_{-\infty}^{\infty} {\mathcal{N}(\mathbf{x}, \mathbf{m}_1, \mathbf{\Sigma}_1)\mathcal{N}(\mathbf{y},  \mathbf{C} \mathbf{x},\mathbf{\Sigma}_2)d\mathbf{x}} \nonumber \\
 &\propto& { \mathcal{N}(\mathbf{y},  \mathbf{C} \mathbf{m}_1, \mathbf{C} \mathbf{\Sigma}_1 \mathbf{C}^T + \mathbf{\Sigma}_2)},
\end{eqnarray}
and
\begin{eqnarray}
\mathcal{N}(\mathbf{x}, \mathbf{m}_1, \mathbf{\Sigma}_1)\mathcal{N}(\mathbf{x}, \mathbf{m}_2, \mathbf{\Sigma}_2) \propto \mathcal{N}(\mathbf{x}, \mathbf{m}_3, \mathbf{\Sigma}_3),
\end{eqnarray}
where the variance is given by
\begin{eqnarray}
\mathbf{\Sigma}_3 = (\mathbf{\Sigma}_1^{-1}+\mathbf{\Sigma}_2^{-1})^{-1},
\end{eqnarray}
and the expectation is given by
\begin{eqnarray}
\mathbf{m}_3 = \mathbf{\Sigma}_3 (\mathbf{\Sigma}_1^{-1}\mathbf{m}_1+\mathbf{\Sigma}_2^{-1}\mathbf{m}_2).
\end{eqnarray}

Below is the notation used throughout the derivation:
\begin{eqnarray}
\pi_{\mathbf{x}_{t-1},\mathbf{x}_{t}}(\mathbf{x}_{t-1}) &=& \mathcal{N}(\mathbf{x}_{t-1}, \mathbf{x}_{\pi_x,t-1}, P_{\pi_x,t-1}) \nonumber\\
\pi_{\mathbf{x}_{t-1},\mathbf{y}_{t-1}}(\mathbf{x}_{t-1}) &=& \mathcal{N}(\mathbf{x}_{t-1}, \mathbf{x}_{\pi_y,t-1}, P_{\pi_y,t-1}) \nonumber\\
\pi_{\mathbf{y}_{t},\mathbf{b}_{t}}(\mathbf{y}_{t}) &=& \mathcal{N}(\mathbf{y}_{t}, \mathbf{y}_{\pi,t}, S_{\pi,t}) \nonumber\\
\pi_{\mathbf{x}_{t}}(\mathbf{x}_{t}) &=& \mathcal{N}(\mathbf{x}_{t}, \mathbf{x}_{l,t}, P_{l,t}) \nonumber\\
\pi_{\mathbf{y}_{t}}(\mathbf{y}_{t}) &=& \mathcal{N}(\mathbf{y}_{t}, \mathbf{y}_{l,t}, S_{l,t}) \nonumber\\
\lambda_{\mathbf{y}_{t}, \mathbf{x}_{t}}(\mathbf{x}_{t}) &=& \mathcal{N}(\mathbf{x}_{t}, \mathbf{x}_{\lambda_y,t}, P_{\lambda_y,t}) \nonumber\\
\lambda_{\mathbf{x}_{t}, \mathbf{x}_{t-1}}(\mathbf{x}_{t-1}) &=& \mathcal{N}(\mathbf{x}_{t-1}, \mathbf{x}_{\lambda_x,t-1}, P_{\lambda_x,t-1}) \nonumber\\
\lambda_{\mathbf{b}_{t}, \mathbf{y}_{t}}(\mathbf{y}_{t}) &=& \mathcal{N}(\mathbf{y}_{t}, \mathbf{y}_{\lambda,t}, S_{\lambda,t}) \nonumber\\
\text{BEL}(\mathbf{x}_{t}) &=& \mathcal{N}(\mathbf{x}_{t}, \mathbf{x}_{\text{BEL},t}, P_{\text{BEL},t}) \nonumber \\
p(\mathbf{x}_{t}|\mathbf{x}_{t-1}) &=&\mathcal{N}(\mathbf{x}_{t}, A\mathbf{x}_{t-1} + B \mathbf{u}_{t-1}, \Sigma_p).\nonumber
\end{eqnarray}

\begin{enumerate}
\item
Step 1: $\mathbf{x}_{t-1}$ \ding{213} $\mathbf{y}_{t-1}$: we have
\begin{eqnarray}
&&\pi_{\mathbf{x}_{t-1}}(x_{t-1})\nonumber\\
&=&\int_{-\infty}^{\infty}{p(\mathbf{x}_{t-1}|\mathbf{x}_{t-2})\pi_{\mathbf{x}_{t-2},\mathbf{x}_{t-1}}(\mathbf{x}_{t-2})d\mathbf{x}_{t-2}} \nonumber \\
&=& \mathcal{N}(\mathbf{x}_{t-1}, \mathbf{x}_{l,t-1}, P_{l,t-1}),
\end{eqnarray}
where the expectation is given by
\begin{eqnarray}
 \mathbf{x}_{l,t-1} &=&  \mathbf{A}\mathbf{x}_{\pi_x,t-2},
\end{eqnarray}
and the variance is given by
\begin{small}
\begin{eqnarray}
P_{l,t-1} &=&  \mathbf{A} \times P_{\pi_x,t-2} \times  \mathbf{A}^T + \mathbf{\Sigma}_p \nonumber \\
\pi_{\mathbf{x}_{t-1}, \mathbf{y}_{t-1}}(\mathbf{x}_{t-1}) &=&  \pi_{\mathbf{x}_{t-1}}(x_{t-1})\lambda_{\mathbf{y}_{t-1}, \mathbf{x}_{t-1}}(\mathbf{x}_{t-1}) \nonumber \\
&=& \mathcal{N}(\mathbf{x}_{t-1}, \mathbf{x}_{l,t-1}, P_{l,t-1})\times 1 \nonumber \\
&=& \mathcal{N}(\mathbf{x}_{t-1}, \mathbf{x}_{\pi_y,t-1}, P_{\pi_y,t-1}) \\
\text{where} \nonumber\\
 \mathbf{x}_{\pi_y,t-1} &=& \mathbf{x}_{l,t-1}; P_{\pi_y,t-1} = P_{l,t-1}.
\end{eqnarray}
\end{small}

\item:
Step 4: $\mathbf{y}_{t-1}$ \ding{213} $\mathbf{x}_{t-1}$: we have
\begin{small}
\begin{eqnarray}
\lambda_{\mathbf{y}_{t-1}, \mathbf{x}_{t-1}}( \mathbf{x}_{t-1}) &=&
\gamma_{\mathbf{y}_{t-1}}(\mathbf{x}_{t-1}) \nonumber \\
 &=& \int_{-\infty}^{\infty}{\lambda_{\mathbf{b}_{t-1}, \mathbf{y}_{t-1}}(\mathbf{y}_{t-1})p(\mathbf{y}_{t-1}|\mathbf{y}_{x-1})} \nonumber \\
&=& \mathcal{N}(\mathbf{x}_{t-1}, \mathbf{x}_{\lambda_y, t-1}, P_{\lambda_y, t-1}),
\end{eqnarray}
\end{small}
where the expectation is given by
\begin{eqnarray}
\mathbf{x}_{\lambda_y, t-1} &=& \mathbf{C}^{-1} \times \mathbf{y}_{\lambda, t-1},
\end{eqnarray}
and the variance is given by
\begin{eqnarray}
 P_{\lambda_y, t-1} &=&  \mathbf{C}^{-1}(S_{\lambda, t-1} + \mathbf{\Sigma}_o) \times ( \mathbf{C}^{-1})^T.
\end{eqnarray}

\item
Step 5: $\mathbf{x}_{t-1}$ \ding{213} $\mathbf{x}_{t}$: we have
\begin{eqnarray}
\pi_{\mathbf{x}_{t-1},\mathbf{x}_{t}}(\mathbf{x}_{t-1}) &=& \pi_{\mathbf{x}_{t-1}}(x_{t-1}) \times \lambda_{\mathbf{y}_{t-1}, \mathbf{x}_{t-1}}(\mathbf{x}_{t-1}) \nonumber \\
&=& \mathcal{N}(\mathbf{x}_{t-1}, \mathbf{x}_{\pi_x, t-1}, P_{\pi_x, t-1}),
\end{eqnarray}
where the variance is given by
\begin{eqnarray}
P_{\pi_x, t-1} &=& ( P_{l,t-1}^{-1} + P_{\lambda, t-1}^{-1})^{-1},
\end{eqnarray}
and the expectation is given by
\begin{eqnarray}
\mathbf{x}_{\pi_x, t-1} &=& P_{\pi_x, t-1} \times ( P_{l,t-1}^{-1} \times \mathbf{x}_{l,t-1} \nonumber \\
&+& P_{\lambda, t-1}^{-1}\mathbf{x}_{\lambda, t-1}). \nonumber
\end{eqnarray}
The belief is thus given by
\begin{eqnarray}
\text{BEL}{\mathbf{x}_{t-1}} &=& \alpha \times 1 \times \lambda_(\mathbf{y}_{t-1}, \mathbf{x}_{t-1})(\mathbf{x}_{t-1}) \nonumber \\
&\times& \pi_{\mathbf{x}_{t-1}}(\mathbf{x}_{t-1}) \nonumber\\
&=& \mathcal{N}(\mathbf{x}_{t-1}, \mathbf{x}_{\text{BEL}, t-1}, P_{\text{BEL}, t-1}),
\end{eqnarray}
where the variance is given by
\begin{eqnarray}
P_{\text{BEL}, t-1} &=& (P_{\lambda_y, t-1}^{-1} + P_{l, t-1}^{-1})^{-1},
\end{eqnarray}
and the expectation is given by
\begin{eqnarray}
\mathbf{x}_{\text{BEL}, t-1} &=& P_{\text{BEL}, t-1} \times (P_{\lambda_y, t-1}^{-1} \times \mathbf{x}_{\lambda_y, t-1} \nonumber \\
&+& P_{l, t-1}^{-1} \times \mathbf{x}_{l, t-1}).
\end{eqnarray}

\item
Step 10: $\mathbf{x}_{t}$ \ding{213} $\mathbf{x}_{t-1}$, we have
\begin{eqnarray}
\lambda_{\mathbf{x}_{t}, \mathbf{x}_{t-1}}( \mathbf{x}_{t-1}) &=& \gamma_{\mathbf{x}_{t}}( \mathbf{x}_{t-1}) \nonumber \\
&=& \int_{-\infty}^{\infty}{\lambda_{\mathbf{y}_t, \mathbf{x}_t}(\mathbf{x}_t)p(\mathbf{x}_{t-1}|\mathbf{y}_{t-1})d\mathbf{x}_{t}} \nonumber \\
&=&  \mathcal{N}(\mathbf{x}_{t-1}, \mathbf{x}_{\lambda_x, t-1}, P_{\lambda_x, t-1}),
\end{eqnarray}
where the variance is given by
\begin{eqnarray}
P_{\lambda_x, t-1} &=&  \mathbf{A}^{-1}(\Sigma_p + P_{\lambda_y, t}) (\mathbf{A}^{-1})^T \nonumber \\
\mathbf{x}_{\lambda_x, t-1} &=& \mathbf{A}^{-1}\mathbf{x}_{\lambda_y, t}.
\end{eqnarray}
\end{enumerate}

\end{document}